# Analysis of Prospective Elements and Crystal Lattice Structures via Computer Algorithms to Identify Standard Temperature Pressure (STP) Superconductors

Ekram M Towsif

Chemistry of Materials and Nanomaterials Course, Chemistry Department, Wesleyan University, Middletown, Connecticut

*Keywords: Superconductivity; Room Temperature; Ambient Pressure; Metastable States; Crystal Structures; Algorithms*

**ABSTRACT:** Superconductors have the potential to revolutionize technology due to their ability to have zero electrical resistance. However, superconductor materials require either low temperatures or high pressures to function in a superconductive state. Thus, researchers are now on the search for the first-ever room temperature, ambient pressure superconductor. Yet, recent discoveries have only shown superconductors that work at low temperatures with ambient pressure or room temperature with high pressures. The region between these two extremes has not been identified due to the number of variables that affect superconductivity. To reduce the number of permutations that need to be tested to identify the first STP superconductor I propose the use of a computer algorithm designed to test various crystal structures of superconducting materials and combinations of elements that will have zero electrical resistance and exhibit the Meisner effect. Some elemental superconductors that have the highest critical temperature at standard pressure are V, Zr, La, Hf, Re, Th, Pa, U, and Am. The elements once combined with other elements in the right crystalline structure, can produce a metastable state where the superconductors will keep their physical characteristics once they form.

## INTRODUCTION:

Superconductivity was first discovered in 1911 when Dutch physicist Heike Kamerlingh Onnes showed the resistance of liquid helium disappeared when cooled down to 4 degrees Kelvin. Another milestone was achieved in 1933 by Walther Meissner and Robert Ochsenfeld for discovering that superconductors produced perfect mirror induced currents of a moving magnetic field. The effect generates diamagnetic behavior where the induced currents are so strong that magnets can levitate above the superconductor, also called the Meissner effect. These two properties became the golden standard to determine if a material is superconducting or not. With these two simple tests, researchers went to work trying to identify other materials beyond simple elemental superconductors such as metallic alloys, organic carbon-based superconductors, and even ceramics. The discovery of various types of superconductors questioned conventional ideas and pushed the boundaries of what was possible. For many years it was long believed that insulators like ceramics do not behave like superconductors or that materials could not be both a superconductor and a natural magnet at the same time. As researchers pushed to uncover the first room-temperature ambient pressure superconductor, the previous notions happened to be proven incorrect, and more discoveries are yet to come (1).

The development of all the new superconductors produced type 1 and 2 classifications, which refer to the number of critical magnetic field strengths that enable superconductive behavior (2). The closest theory that explains how superconductors work is known as the BCS theory. The theory states that the above superconductor effects arise from the condensation of Cooper pairs, which are a pair of electrons weekly joined together at a critical temperature, magnetic field, or pressure such that they are less likely to be affected by electron scattering due to impurities, defects, and vibrations in the crystalline structure of the material. This theory completely works for low-temperature type 1 superconductors, but for high-temperature type 2 superconductors, BCS is inadequate and cannot fully explain the behavior. However, we can still use the mathematical correlations from the theory and with new equations relating to superconducting critical temperature to understand the phenomenon better before trying to achieve STP superconductivity (1).

Nonetheless, not understanding the physical nature that gives rise to these behaviors does not impose restrictions on utilizing their potential. There are enormous beneficial applications of superconductors, such as high-efficiency energy transfer of electricity through superconductive electrical cables, transportation via maglev trains, improved MRI machinery, NMR technology, particle accelerators, fusion reactors, and more (3). The beneficial applications for such materials are enormous, and waiting to come up with a more well defined mathematical theory on how superconductivity works may take decades. Even then, the materials that will be ideal for each application mentioned above will most likely differ, and the theory will not be able to predict which combinations of materials will be optimal in each situation. The two desired characteristics that researchers and companies want to utilize, the Meissner effect and zero electrical resistance, were found a hundred years ago. However, superconducting materials used today require low temperatures or high pressures. Achieving such conditions requires exposing the materials to either liquid nitrogen/helium or diamond anvils, respectively. Subsequently, there needs to be a more holistic approach in identifying superconducting materials, possibly utilizing an optimized feedback loop with both theory and synthesis.

## HYPOTHESIS:

Thus, the race to find the material that behaves like a superconductor at standard chemical conditions has begun. For superconducting materials, this refers to a pressure of 1 atmosphere and a temperature of 298 Kelvin. I hypothesize that the first STP superconductors are composite materials where the addition of specific elemental compounds assists in stabilizing the structure at a critical temperature of 298 Kelvin and pressure of 1 atm at a specified magnetic field.

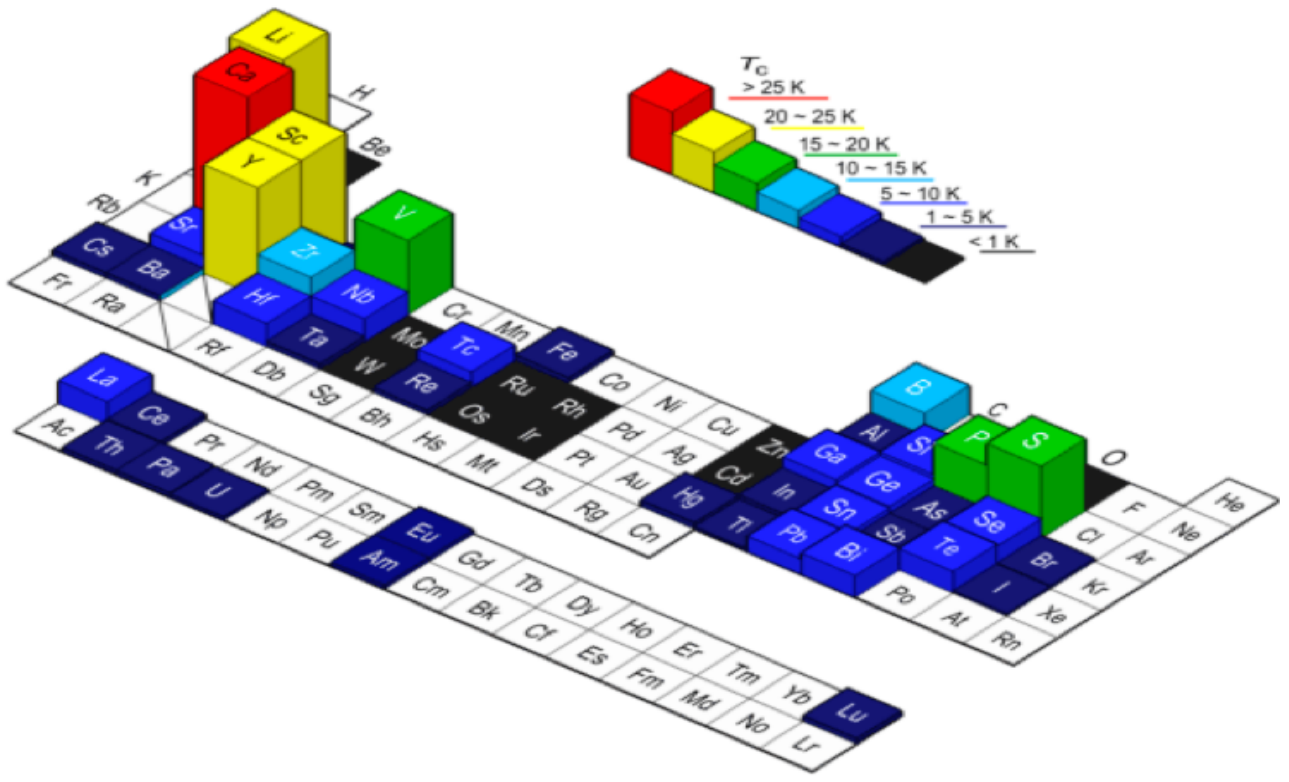

*Figure 1) Critical temperatures of elemental superconductors (4).*

The specific elements that would stabilize composite superconductors such that they have a higher transitional temperature can possibly be the natural elemental superconductors that have the high transitional temperatures. From figure four, the elements with the highest transition temperature are calcium, lithium, yttrium, and scandium. Thus, adding these elements to composite materials will result in higher transitional temperatures for the superconductive state. Experiments have shown an increase in transitional temperature for copper oxide materials. For example, yttrium barium copper oxide (YBCO) is a type two superconductor that is easy to make and has one of the highest critical temperatures to date of 92 K. Yet, YBCO is still not the material closest to STP superconductivity. Mercury barium calcium copper oxide (HBCCO) has the highest cuprate critical temperature of 135 K, but HBCCO is hard to synthesize (5). Likewise, elements can reach a superconducting state under high pressures such as yttrium and calcium. The increased pressure pushes the critical transition temperatures in the materials higher. Recent developments in high-pressure experiments with materials have revealed composites, carbonaceous sulfur hydride (CSH8), that are superconducting at temperatures around 288.15 K at pressures of 260 GPa (6,7).

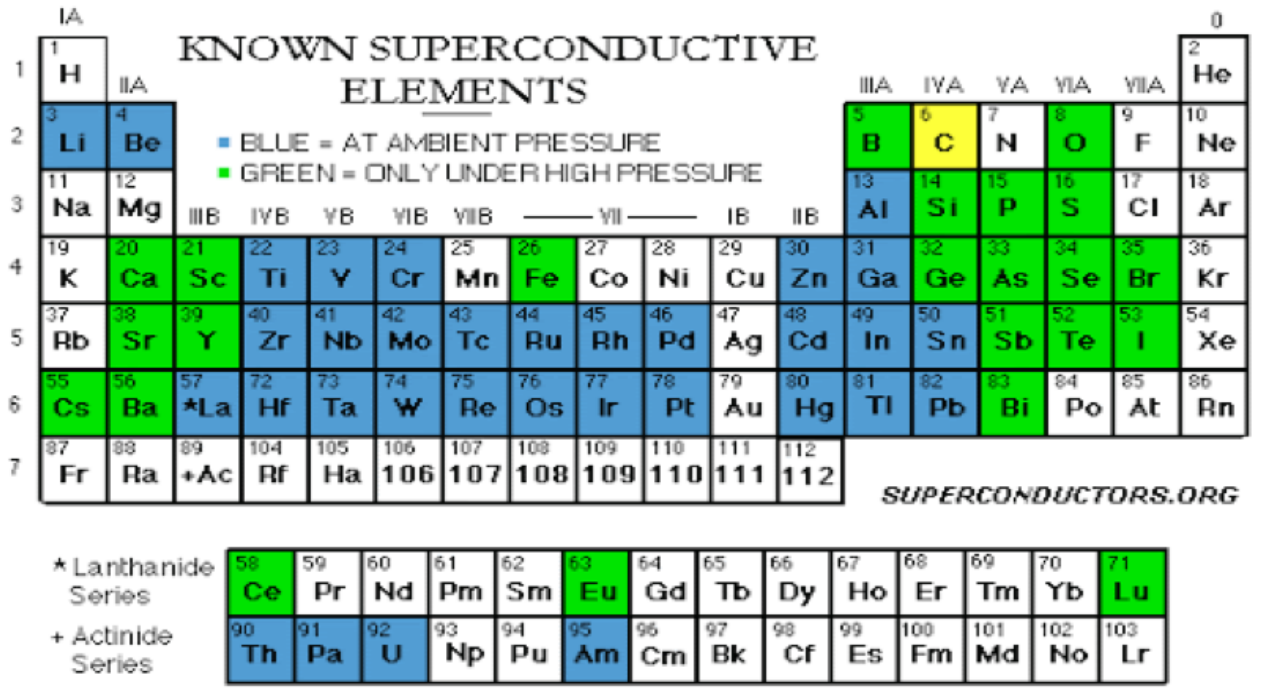

*Figure 2) Pressure diagram of elemental superconductors (2)*

Therefore, to overcome the temperature and pressure barriers, figures 1 and 2 have to get cross-checked to identify known elemental superconductors with the highest transitional temperatures under ambient pressure. Reviewing the two figures, the nine elements that have the potential for stabilizing the physical properties of composites are; vanadium, zirconium, lanthanum, hafnium, rhenium, thorium, protactinium, uranium, and americium. Synthesizing new types of superconductors with these elements may yield even high transitional temperatures at lower pressures. Using the HBCCO (cuprate) and the CSH8 (hydride) as templates and interchanging the nine substances above may result in the first-ever STP superconductor. In fact, researchers have already designed a compound, LH10, with a critical temperature of 250 K at 170 GPa, which supports the predictions that the nine elements may increase the transition temperature and or reduce the necessary pressure. Thus, the cuprate and hydrides backbones, examined with the addition of one of the nine elements, has potential (8).

Consequently, the fine-tuning of the superconductor composite for room temperature and ambient pressure conditions will enable us to understand the thermodynamic reaction occurring during the transition from a superconductive state to a non-superconductive state. Understanding the energy transitions may help develop metastable crystalline structures, which may slow down the reversal of the superconductive state as temperatures increase or pressures decrease. Eventually, forcing the superconductors to become stable as it is gradually brought back up to either room temperature or brought down to standard pressure, depending on the starting synthesis method used. A similar metastable structure is achieved naturally with carbon lattices, particularly during the formation of diamonds. Diamonds are naturally occurring gems that are considered to be one of the most durable materials on the planet and are used to create the immense pressures needed to make high-pressure room temperature superconductors in diamond anvils. However, at STP, the stable form of carbon is graphite, but diamonds never transform into graphite unless at low pressures and high temperatures. Thus, changes in the superconductive states get produced by incorporating new elements into known superconductors to allow isomerization to occur in the crystalline structure changing the critical values. Diamonds also take billions of years to form under the Earth before they reach a metastable state. Therefore, there may be an added time component missing necessary to allow superconductors to develop a stable crystalline structure at room temperature and pressure (9).

## METHODOLOGY/PREDICTED RESULTS:

Regardless, there are too many variables for researchers to test within the laboratory, which means there needs to be a way to tests specific ranges for the temperature, pressure, time, and other parameters for the crystalline lattice structures that enable us to get the closest room temperature and pressure. Ultimately, simulating the superconductive behavior for various compositions of elements in different structural forms under controllable temperature and pressure conditions over some time is an ideal method. Due to the computationally intensive nature of such a simulation, resulting from all the dynamical variables, the utilization of supercomputers may be required alongside machine learning algorithms (10). A training data set, superconductor data obtained from SuperCon or other databases, of all known superconductors and their properties to date, can be inputted into the machine learning neural network. The machine learning neural networks become implemented in various ways, and one such method utilizes the data to make the program predict the critical temperature value, Tc, based on the properties of each material. The predictions on the Tc value get cross-checked with the actual experimental values through a random forest regression model. The model estimates the value of the correlation coefficient (r), which measures the strength of the linear relationship. Once the model approximates the r value to be above 0.90, then the model is good enough to predict the Tc values of other unknown superconductors. Before the recent discovery of the CSH8, the model could not predict Tc values more than the value in the training data set. Currently, 288.15 K is the highest Tc value that can be implemented in the model. After the model identifies Tc values greater than 273.15 K, composite materials can get synthesized to confirm the validity of the model. If the results get verified, the model continues to explore more possible structures, and if the results are inaccurate, then the model gets retrained on more data to correct the previous predictions. Ternary structural predictions, shown in figure 3, demonstrates how the regression identifies Tc values and maps them. The figure illustrates a triangle with three different elements at the vertices. Each point on the perimeter and area of the triangle represents a possible superconductor with the form. Each combination has a Tc value shown through the heat map of the triangle. The regression model can devolve various ternary structural heat maps with different combinations of elements. Moreover, the model can expand by training on more data to study the heat maps of higher-order polynomials in the form of cuprates or hydrides. The predictions of the Tc value indicates the region of zero resistance in the materials (11).

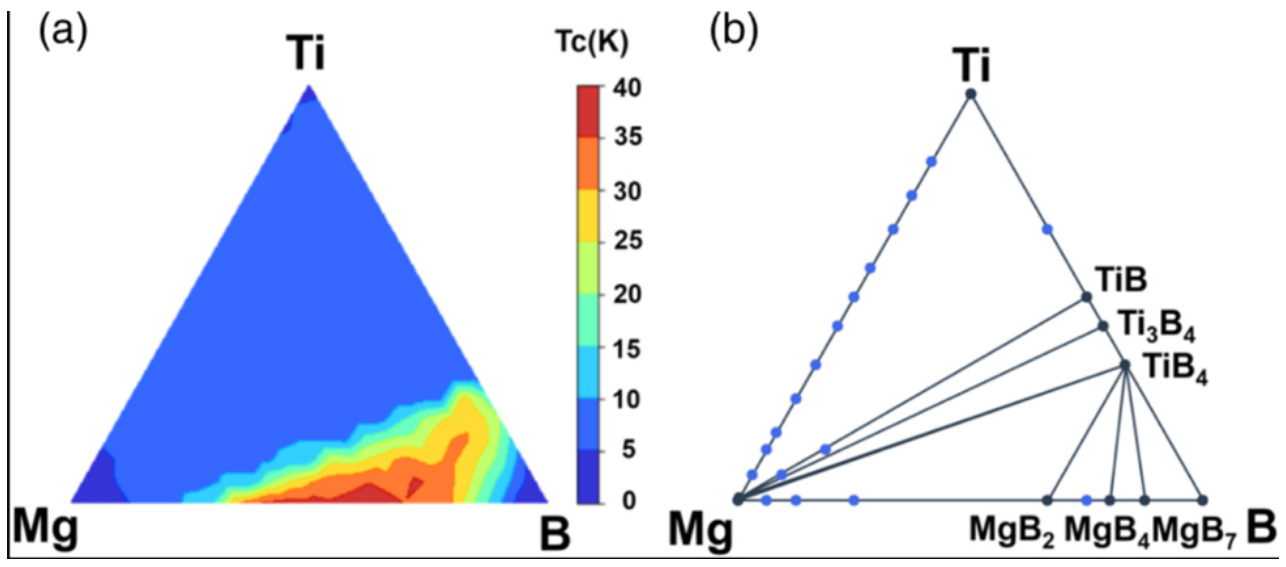

*Figure 3) Machine Learning Simulations to find critical temperature of ternary superconductors (11).*

The model applied with the copper oxide, hydrogen, and the nine elements identified above at each vertex of the triangle can create a new superconductor that exhibits hybrid behaviors of cuprates and hydrides. The new "cuprate hydride material" can be stabilized with one of the nine materials. Since cuprates are the best ambient pressure high-temperature superconductors and hydrides are known to be the best room temperature high-pressure superconductors, combining them may result in the best of both materials; an STP superconductor. Similarly, an altered version of the model can predict the critical magnetic field (Bc) where these type two composite superconductors will exhibit the Meissner effect (12).

Thus, every possible iteration can be done by the computer simulation with the information provided. The testing parameters, the Meissner effect, and zero resistance used to identify the creation of a superconductor have been computationally implemented and experimentally verified. Once the simulation runs with minimal errors, a variety of potential new superconducting materials will come into existence. The computer would be able to log the composition and structures of such materials as well as the critical temperature and pressure. Likewise, with additional modifications of the machine learning algorithm, it will also record how long the superconducting state will last due to structural and elemental composition differences after the temperature and pressure reset to STP. All of the information that the computer simulation provides hints for finding a synthesis method for the ideal candidates. The use of X-ray crystallography techniques to verify the crystalline structure of the synthesized materials can help reduce defects present. Upon synthesizing, cross-checking the new superconducting material with the simulation provides the accuracy of the experiment enabling improvements for further simulations (13).

Although the simulation can guess the ideal combination of elements that have a Tc value obtainable at STP, there can still be a structural component that may need to be simulated as well. The arrangement of the elements can have effects on the Tc value, indicating that the crystalline lattice affects how the cooper pairs in the materials form and interact. Thus, identify the relationship between the transition temperature and crystal structure of superconductor metals and alloys can shed light on the optimal lattice for cooper pairs to form at STP conditions. The BCS theory states that the cooper pairs form due to minimal scattering of the electrons from the lattice resulting in the ability of the cooper pairs to move at the same speed, direction, and in phase with each other. The Roeser–Huber equation correlates the geometric structure of crystals to their critical temperatures. The equation uses the Planck's, the Boltzmann's, and the electron mass constants, while the mass of the charge carriers ML, the doping distance of the crystal x, the number of superconducting layers n, and the type of elements n1/n2 vary. Using the equation with various parameters produces a linear relationship with the critical temperature, as shown in figure 4. More interestingly, an array of superconducting materials fall on the linear fit with small deviations from the predicted values. Thus, using the Roeser–Huber equation in conjunction with the optimal superconductors the machine learning algorithm provides, iterations of crystal structures can run. Each iteration will indicate how much the Tc value of the newly identified superconductor will respond to structural changes. With the composition and structural data, the computer can run an optimized synthesis of the ideal superconducting material (14).

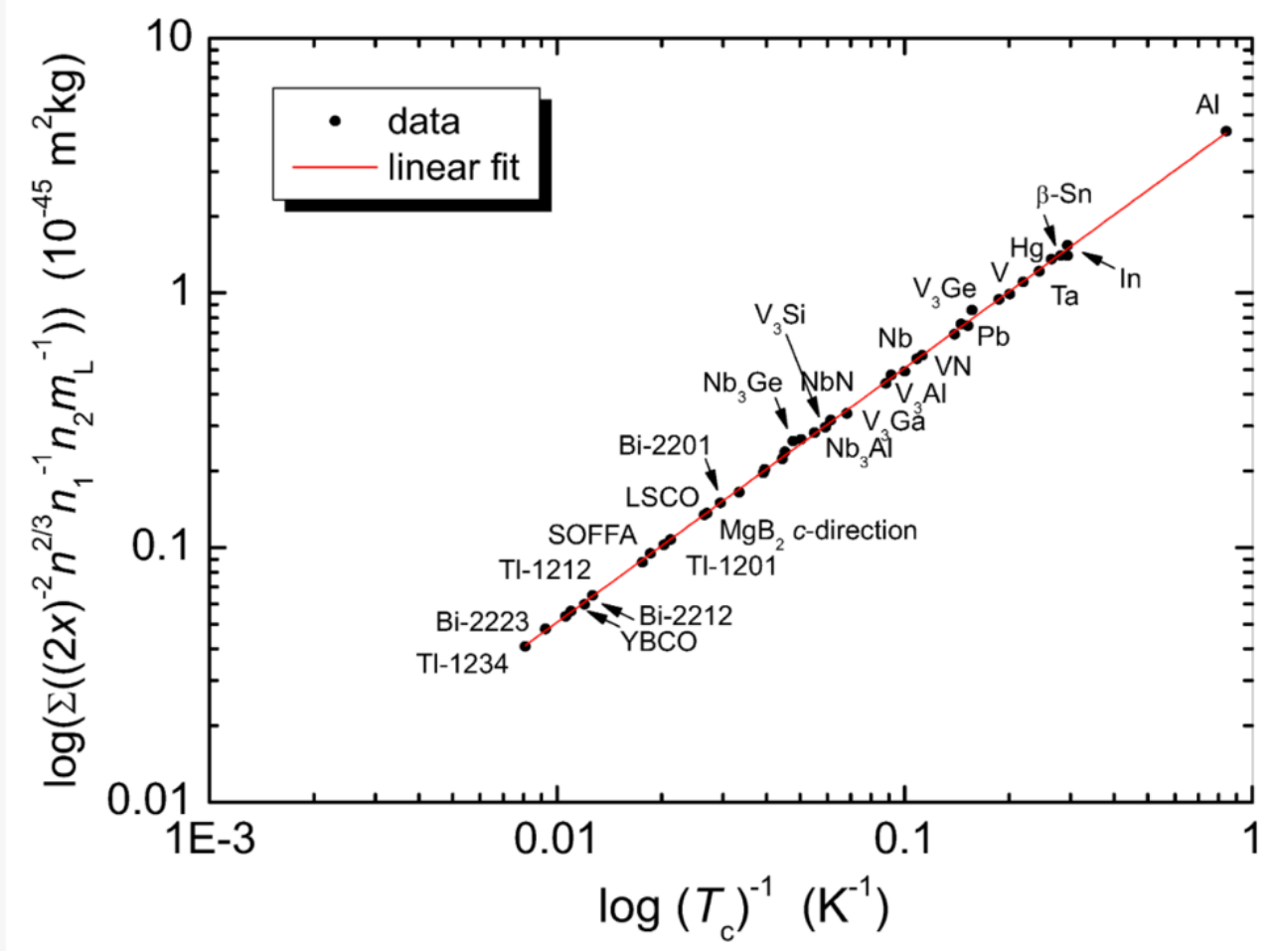

*Figure 4) Log scale graph correlating simulated crystal structures of superconductors and their critical temperature (14).*

## CONCLUSION:

In conclusion, superconductivity has developed experimentally to the point where research is inching closer to the ideal STP superconductor. However, to accelerate the advancements, I propose the use of machine learning algorithms to test a novel hydride material composed of cuprates, hydrides, and one of nine promising elements. The various compositions can be searched through the program, resulting in predicted Tc values. The materials with the highest values will have the Roeser–Huber equation be applied to identify the optimal structural form with another predicted Tc value. If the Tc values are reasonably similar, then experimental synthesis of the superconductor can get carried out to verify the Tc value and improve the model with more data. After various iterations of the process, the model will improve exponentially, eventually resulting in the identification of the first-ever STP superconductor.

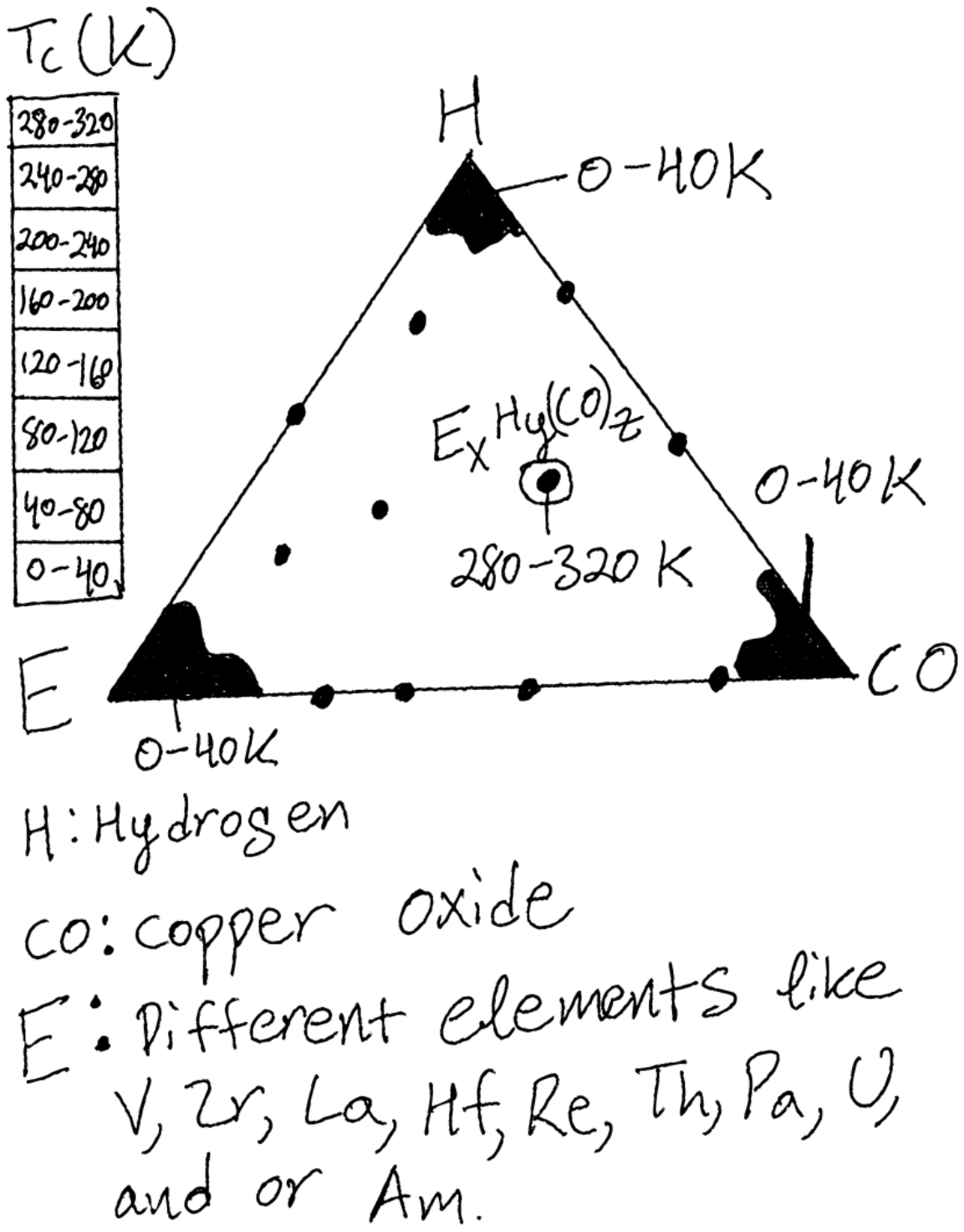

**Picture Summary**: With the recent development of room temperature high-pressure superconductors, new critical temperature (Tc) data can be used as training data for a computational regression analysis model. We propose a computational research approach to finding standard temperature and pressure (STP) superconductors. I anticipate that running the analysis on various stoichiometric ratios of a hybrid elemental hydride cuprate will facilitate in predicting the Tc values, providing more insights into a potential synthesis for STP superconductors. Furthermore, I am anticipating that the program will identify stoichiometric ratios with a Tc value in the range of 273 – 300 K, and only by experimentally testing will we know for sure if the analysis worked. However, if it did not, then the data will get fed back to retrain the model, exponentially making it better for future tests ending with the discovery of the first-ever STP superconductor.